# *Waveguide-integrated plasmonic photodetectors and activation function units with phase change materials*

Jacek Gosciniak

*Independent Researcher, 90-132 Lodz, Poland*
*Email: jeckug10@yahoo.com.sg*

**Abstract**
With a rapidly growing amount of data generated and processed, a search for more efficient components and architectures such as neuromorphic computing that are able to perform a more and more complex operations in more efficient way continue.
Here we show that thin films of chalcogenide phase change materials (semiconductors) can serve as building blocks for novel type of plasmonic components that operate seamlessly in both electrical and optical domains without the need for repeated electrical-to-optical conversions.
In consequence, novel waveguide-integrated devices were proposed that are able to operate simultaneously as photodetectors and activation function units and that are based on low-loss plasmonic waveguide platform and phase change materials. Theoretically predicted coupling efficiency exceeding 95 % and extremally low insertion losses of 0.01 dB/µm in connection with an enhanced light-matter interaction provided by plasmonics enables a realization of very efficient and compact photodetectors and activation function units. A detection and threshold mechanism involve a Joule heating of phase change materials through internal or external metal contacts. With this paper, different arrangements and operation conditions were analyzed to ensure most efficient on-chip signal processing.

**Introduction**
In the last years, photonic integrated circuits (PICs) became very attractive as they offer broad bandwidth and very efficient information transport, processing, and storage [1, 2]. Large-scale PICs are based on the silicon photonics platform that is compatible with the well-established CMOS technology.
Two key advantages of integrated photonics over its electronic counterparts rely on the massively parallel data transfer in conjunction with multichannel sources and extremely high data modulation speed limited only by the bandwidth of on-chip optical modulators and photodetectors. Thus, modulators and photodetectors are very essential components in the network [3, 4]. As modulators perform electrical to optical signal conversion [5, 6, 7, 8, 9], the on-chip photodetectors convert light into an electrical signal [10, 11, 12, 13]. Being the last components in the optical links, the photodetector must operate with low power if costly amplifiers are to be avoided.
Photonic integrated components can play an important role in a newly established field in the area of neuromorphic platform, so called the neuromorphic photonics, that assume to mimic a behavior of human brain through photonics [14, 15, 16, 17] or plasmonics [18]. To fully develop this concept, two classes of devices need to be engineered - the weighted sum/addition and the nonlinear activation [16, 17]. The weighted sum/addition can be obtained by a combination of photodetectors, modulators and switches where the optical carriers or modulated signals are attenuated or delayed by certain amount dictated by the weight values. Then, a nonlinear unit applies an activation function *f* to the weighted sum/addition, yielding the output of the neuron that is transmitted to many other neurons. The activation function mimics the firing feature of biological neurons. Thus, the nonlinear activation function can be used to set a threshold from which to define activated and deactivated behavior in artificial neurons [14].
As most traditional photodetectors and activation functions suffer either from high power consumption, low efficiency, low operation speed or large footprint, the progress in development of such devices needs to be made even through a new material platform [19, 20, 21], a novel design that is based on plasmonics [22], or both.

**Plasmonics platform - LR-DLSPP arrangement**
Plasmonics can squeeze light much below the diffraction limit, which reduces the device footprint [22]. Furthermore, a small device volume means a higher density of integration and, simultaneously, lower power consumption, easier heat dissipation, and faster operation speed [23]. Taking those advantages, plasmonics can serve as a platform for building novel on-chip photodetectors with improved performances [10, 11, 12, 13]. However, as the optical energy received at a photodetector is directly related to transmitter optical output power and the total link loss power budget, thus minimizing the attenuation losses and coupling losses at photodetector is crucial for overall performance of the system [10]. High coupling efficiency means that most of the power coupled to the photodetector can participate in an electrical signal generation.

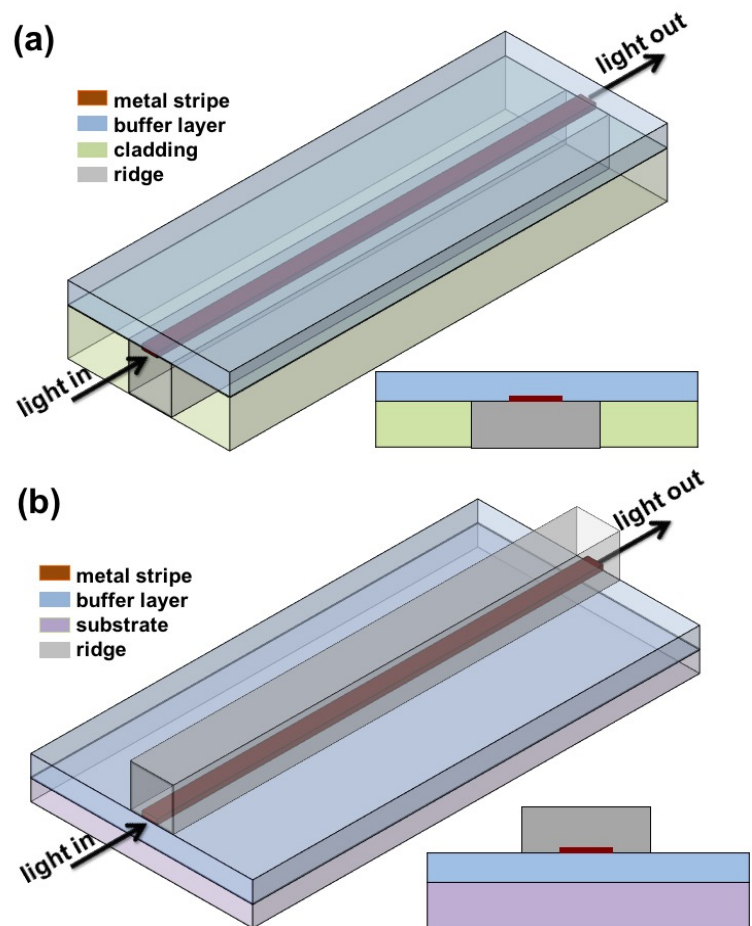

**Figure 1a.** Proposed LR-DLSPP waveguide configuration in (a) "inverse" design and (b) "normal" design.

The proposed photodetectors are based on the long-range dielectric-loaded surface plasmon polariton (LR-DLSPP) waveguide arrangement with the metal stripe placed between the ridge and buffer layer that support TM-polarized mode (Fig. 1a and 1b). Thus, the metal stripe is an essential part of the LR-DLSPP waveguide and can serve as one of the electrodes [24, 25, 26, 27].
In a balance conditions, the mode effective index below a metal stripe is close to the mode effective index above a metal stripe. In consequence, the absorption in metal stripe is minimalized and the propagation length is enhanced. When the active material is deposited either into a buffer layer in "inverse" design (Fig. 1a) or into a ridge in "normal" design (Fig. 1b) even the small change in the refractive index of the active material can disturb a balance.

As a result, the absorption in metal stripe arises and propagation length of the mode decreases. This effect can be highly enhanced when the active material is placed directly at the contact with a metal stripe, *i.e.*, in the electric field maximum of the propagating mode.

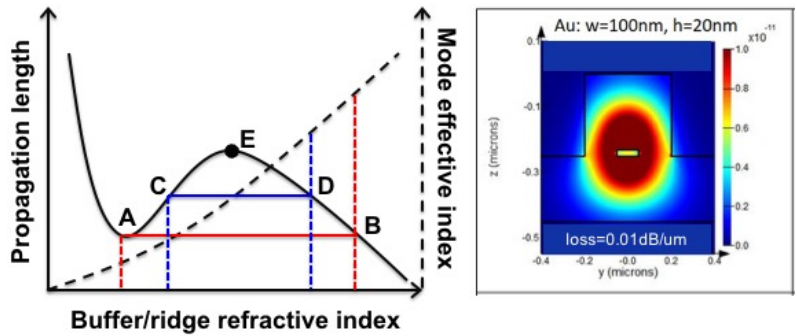

**Figure 2.** (a) Propagation length (solid line) and mode effective index (dashed line) as a function of buffer layer / ridge refractive index and (b) mode effective index and calculated losses for LR-DLSPP waveguide with Au stripe.

The proposed photodetector ensures low absorption losses below 0.01 dB/μm that can be achieved with Si platform (Fig. 2b). Consequently, the insertion losses (IL) below 0.1 dB can be obtained for 10 μm long active region of the photodetector. Even farther reduction in absorption losses can be achieved with lower index waveguide materials such as for example SiN, where the propagation length of 700 μm was measured at telecom wavelength for CMOS-compatible silicon nitride (SiN) [27].

Furthermore, the proposed photodetector ensures high coupling efficiency with the photonic platform that was numerically estimated at 97 %. [24, 25]. Thus, the coupling losses per interface as low as 0,05 dB can be achieved. It was experimentally validated, where the coupling efficiency exceeding 75 % per interface was achieved. A difference between numerical calculations and experimental results was attributed to the presents of thin layer of titanium used for improving adhesion between gold stripe and substrate that introduces substantial mode absorption [26].

In consequence, apart from a complex design and efficient conversion mechanism, the proposed photodetector can provide low attenuation losses and very high coupling efficiency what can highly improve the overall performance of the system through an enhanced electrical signal generation.

**Materials platform - phase change materials (PCMs)**
Recently, the phase change materials (PCMs) have been proposed as very promising materials for realization of nonvolatile optical modulators, switches [19, 20, 28, 29, 30, 31] and photodetectors [32, 33].

They provide extremely high refractive index contrast ($\Delta n$=0.6÷3.0), ultrafast transition (>1 ns), energy-efficient reversible switching, nonvolatility (leading to zero-static power consumption), high scalability, long-term retention (<10 years), and high cyclability ($10^{12}$ switching cycles) [34, 35, 36, 37]. Mechanism of switching in the PCM is realized through a phase transition between its two different phases – amorphous and crystalline [35, 36]. The amorphous state is characterized by high electrical resistance while the crystalline state by low electrical resistance. In terms of the photonics, the amorphous state shows a high transmission while the crystalline state a low transmission as a result of higher imaginary part of the complex refractive index [23].

Switching between different phases (states) of PCM can be performed by Joule heating of the PCM using external heaters (thermal effect) [38, 39], electrical pulses (electro-thermal effect) [23, 28, 40], or optical pulses (photo-thermal effect) [19, 23, 31]. In photonic nonvolatile phase-change switches, the heat-induced refractive index change of PCM induces change in transmitted light.

Mechanism of the phase transition between amorphous and crystalline states in the PCM occurs at a crystallization temperature, $T_c$. Thus, below $T_c$ temperature, the PCM is in the amorphous state. When temperature of the PCM exceeds the $T_c$ the PCM undergoes transition to the crystalline state in which it remains even after cooled back to temperature below $T_c$. To move back to the amorphous state, the PCM needs to be first heated to the temperature above its melting point $T_m$ and then rapidly cooling back to temperature below $T_c$.

$Ge_2Sb_2Te_5$ (GST) and $Ge_2Sb_2Se_4Te_1$ (GSST) are the most commonly used PCMs for integrated phase-change devices owing to its wide availability, well established technology, relatively low switching temperature and very large refractive index shift [19, 35, 39, 40]. The crystallization temperature of GST and GSST was measured at 160 °C while the melting temperature at 600 °C with a quenching time in the range of nanoseconds [41]. In terms of a switching time between states, it is generally slower for crystallization and physically limited to hundreds of picoseconds [20, 42].

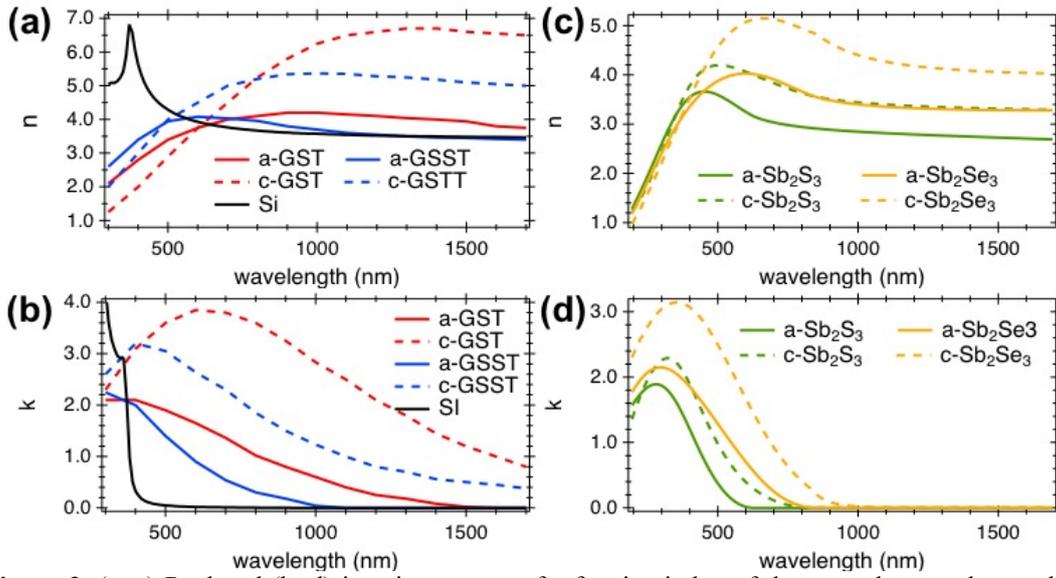

**Figure 3.** (a, c) Real and (b, d) imaginary parts of refractive index of the amorphous and crystalline phased of $Ge_2Sb_2Te_5$ (GST), $Ge_2Sb_2Se_4Te_1$ (GSST) [19], $Sb_2S_3$, and $Sb_2Se_3$ [21] from the visible range to near-infrared. The results were compared with Si.

The refractive index of GST at a wavelength of 1550 nm at amorphous state (a-GST) is $3.80+0.025i$ while at crystalline state (c-GST) is $6.63+1.09i$ (Fig. 3). This provides very large refractive index shift $\Delta n=2.74$ that can be achieved by switching between phases. In comparison, the refractive index of GSST at amorphous state (a-GSST) is $3.47+0.0002i$ while for crystalline state (c-GSTT) it grows to $5.50+0.42i$ what provides a refractive index shift $\Delta n=2.03$. As observed, the refractive index of both a-GST and a-GSST is very close to the refractive index of Si, $n=3.47$ (Fig. 3), what provides a good mode matching. However, both PCMs show large absorption losses in the crystalline phase with $k=1.09$ and $k=0.42$ for c-GST and c-GSST, respectively. As a result, a transmitted light through GST- or GSST-based waveguides undergo much larger change in amplitude than in phase during a phase transition. This limits their use to amplitude modulators rather than to phase modulators. Recently, a new class of low loss PCMs such as, for example $Sb_2S_3$ and $Sb_2Se_3$ were demonstrated that can be considered as reversible alternatives to the standard commercially available chalcogenide GST and GSST [21]. At a wavelength of 1550 nm, $Sb_2S_3$ shows a complex refractive index $n=2.71+0i$ for amorphous phase (a-SbS) and $n=3.31+0i$ for crystalline state (c-SbS) while for $Sb_2Se_3$ it is $n=3.28+0i$ and $n=4.05+0i$ for amorphous (a-SbSe) and crystalline (c-SbSe) phases, respectively. Thus, a contrast of refractive index of $\Delta n=0.60$ for $Sb_2S_3$ and $\Delta n=0.77$ for $Sb_2Se_3$ is achieved while maintaining very low losses, $k<10^{-5}$ (Fig. 3).

**Photodetector arrangement = plasmonics + phase change materials**

Taking into a consideration aforementioned advantages of the LR-DLSPP plasmonic waveguide and PCMs, a novel plasmonic PCM-based photodetector was proposed. The proposed plasmonic photodetector (Fig. 4, 5 and 6) enables reversible switching the states of PCM between its amorphous (high resistance, high transmission) and crystalline (low resistance, low transmission) by sending optical pulses through the waveguide. Furthermore, the state of PCM can be switched by Joule heating using one of the metal electrodes that can work as a heater.

The metal stripe is an essential part of the LR-DLSPP waveguide and can be implemented as an internal heater [24, 25, 26, 27]. As it is in direct contact with a PCM, the heating process can be very efficient. Furthermore, the metal stripe is placed in the electric field maximum of the propagating mode, thus even small change in a refractive index of PCM close to the metal stripe can highly influence the propagating mode. The metal stripe in the proposed photodetector can be limited in length to the active length of the photodetector or can exceed it. As the absorption losses in metal are related to the length of the metal stripe, the absorption losses in metal can be reduced by decreasing a length of the metal stripe.

The proposed LR-DLSPP-based photodetector can be arranged with the external electrodes (contacts) as shown in Fig. 4, 5 and 6. They can be arranged in two different configurations: lateral (Fig. 4) and vertical (Fig. 5, 6). The lateral configuration consists of two metal electrodes placed directly in contact with the PCM (Fig. 4) that is part of the waveguide, thus providing more heat to the PCM locally what lowers the switching threshold. In comparison, the vertical configuration consists of metal electrode placed on top of the either buffer layer (Fig. 5) or ridge (Fig. 6). Depending of the requirements, the electrode can be separated from PCM layer through very thin conductive layer to minimize the mode-overlap with the metal and eventual scattering while still providing an efficient heat transfer to the PCM.

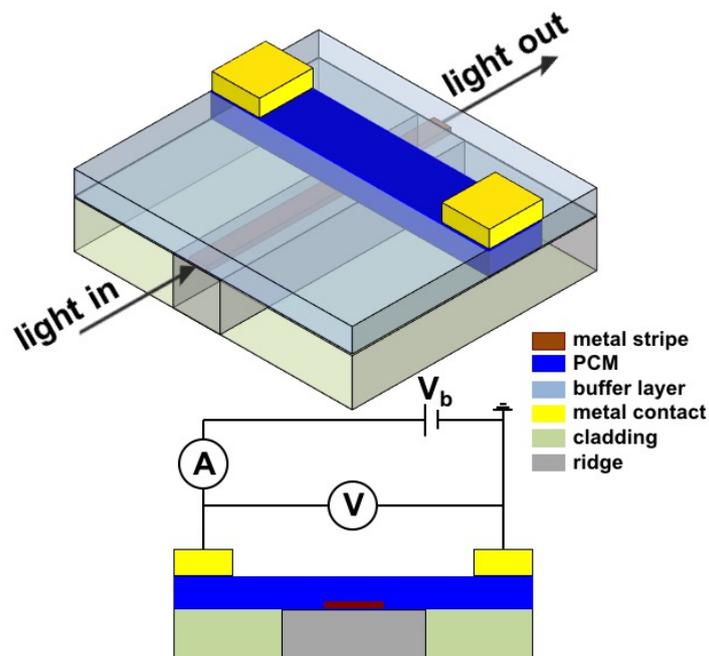

**Figure 4.** The PCM-based photodetector or/and activation unit in „inverse" design with lateral electrodes arrangements.

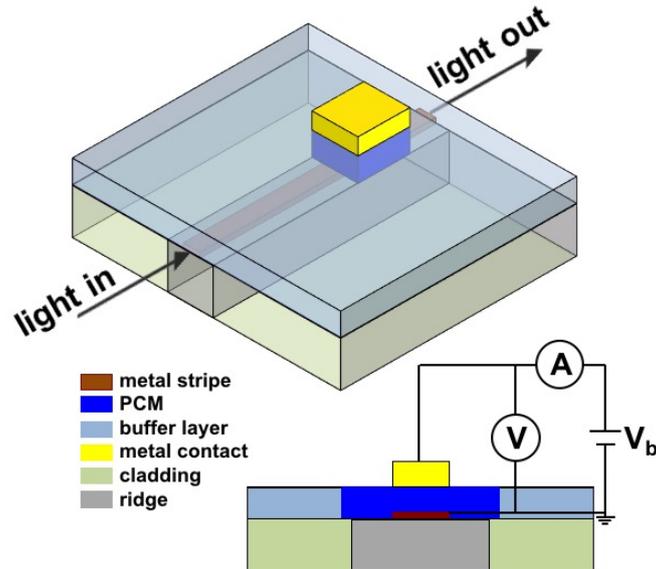

**Figure 5.** The PCM-based photodetector or/and activation unit in „inverse" design with vertical electrodes arrangement.

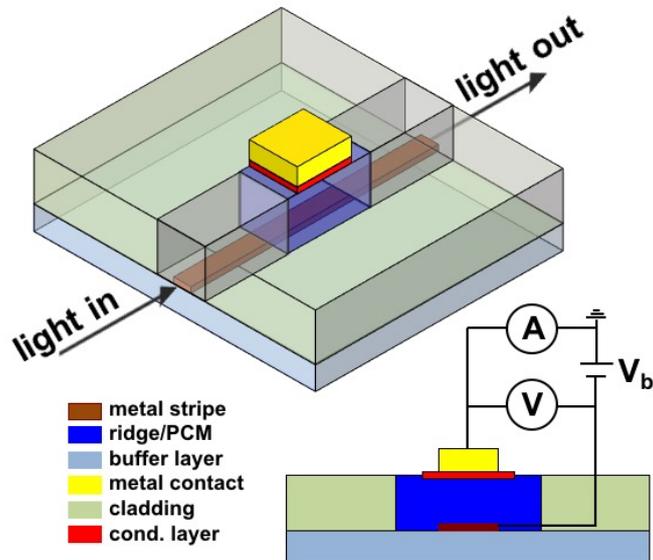

**Figure 6.** The PCM-based photodetector or/and activation unit in „normal" design with vertical electrodes arrangement.

**Heating mechanism**

Under an applied voltage to the metal stripe/electrode, the electrical energy is dissipated into heat. The heat from a metal stripe dissipates to any materials that are in contact with the metal stripe through conductive heat transfer [43]. The amount of heat transfer to the area of interest (ridge or buffer layer) depends upon the thermal conductivity coefficients of the ridge and buffer layer, contact area and thickness of the ridge and buffer layer. Thus, to ensure an efficient heat transfer to the PCM, the second material that constitutes for LR-DLSPP waveguide should possess low thermal conductivity coefficient [43].

The optical power absorbed by the metal stripe is dissipated into any materials that are in contact with a metal stripe and an amount of heat dissipated into PCM depends on thermal resistance and capacity of the surroundings materials. The dissipated power by metal stripe depends on the SPP attenuation coefficient and the length of the active region. Thus, the higher attenuation coefficient, the higher increase of the metal stripe temperature. As it has been previously shown, the temperature increases of the materials that are in contact with a metal stripe is proportional to the power coupled to plasmonic waveguide [44].

The thermal resistance of an object is defined as $R_{th}=L/(\kappa \cdot A)$ and it describes the temperature difference that will cause the heat power of 1 Watt to flow between the object and its surroundings. In comparison, the thermal capacitance of an object, $C_{th}=C_p \cdot \rho \cdot V$, describes the energy required to change its temperature by 1 K, if no heat is exchanged with its surroundings [43]. Here, L is the length of substrate along the heat transfer direction, A is the cross-section area of the substrate, V is the heated volume of the material, $\kappa$ is the thermal conductivity of the material, $C_p$ is the specific heat and $\rho$ is the mass density.

**Table 1.** Thermal properties of the materials constituting the device.

| Material | Refractive index n (1550 nm) | Thermal cond. coeff. $\kappa$ (W/mK) | Heat capacity $C_P$ (J/gK) | Density $\rho$ (g/cm$^3$) |
|---|---|---|---|---|
| SiO$_2$ | 1.45 | 1.4 | 1.4 | 2.19 |
| Si | 3.47 | 148 | 0.713 | 2.32 |
| Au | 0.55+11.5i | 318 | 0.13 | 19.32 |
| Air | 1 | 0.026 | 1.005 | - |
| a-GST | 3.89+0.025i | 0.19 | 0.213 | 5.87 |
| c-GST | 6.63+1.089i | 0.57 | 0.199 | 6.27 |

Thus, materials with lower thermal conductivity coefficient are characterized by higher thermal resistance, so lower electrical powers are required to increase a material temperature as the heat loss is reduced. Simultaneously, materials with lower product of a specific heat and a mass density requires less energy delivered to the material to change its temperature by 1K. Thus, the lower thermal conductivity, the lower heat loss while the lower specific heat and mass density, the lower energy required to heat a material.

As observed from Table 1, a GST possess lower thermal conductivity coefficient and heat capacity than Si and SiO$_2$. In consequence, less power is required to heat a GST what makes a heating process very efficient.

As the proposed LR-DLSPP plasmonic mode is tightly bounded to the metal stripe, the external electrodes can be placed very close to the propagating mode without influencing the propagation length. Thus, the ohmic losses due to the presence of external metals can be minimalized. As it has been previously shown for a lateral configuration [11], the external electrodes can be placed as close as 100 nm away from the ridge without introducing any additional losses [11]. For comparison, an essential increase of additional losses was observed for a similar arrangement but realized with Si photonic waveguide were presence of external electrodes contributed to 0.11 dB/µm additional losses for propagating TM mode [39].

Direct contact of PCM with the electrodes in the proposed photodetector allows lowering the threshold voltage for delivering the right amount of heat for inducing a phase transition in the PCM. A resistive heater optimized for efficient phase transition and contemporary not generating insertion losses can be made in doped silicon or in silicide positioned next to the waveguide [30]. Furthermore, the external heater can be made from transparent conductive oxides (TCOs) such as for example ITO [38, 39] that is characterized by low optical losses at 1550 nm [8, 45]. Apart from it, graphene can be very efficient platform for realization of such a task [28, 29].

The proposed photodetector arrangement allows the electrical readout through resistance measurements. The state of PCM that is a part of the waveguide can be reversibly switched between amorphous and crystalline state by sending a light (optical pulses) through the LR-DLSPP waveguide (Fig. 4, 5 and 6). The guiding properties of the LR-DLSPP waveguide depend on the state of the PCM. For an amorphous state of PCM, the LR-DLSPP waveguide works in a low-loss regime as a mode effective index below a metal stripe is close to the mode effective index above a metal stripe. Thus, the LR-DLSPP waveguide is in balance.

Under a heating of PCM through the light (optical pulses), the PCM undergoes a transition to a crystalline state that is characterized by much higher refractive index (Fig. 3). Thus, the balance in mode effective indices is broken as one side of the LR-DLSPP waveguide (upper or lower) with PCM shows higher mode effective index. In consequence, most of the optical energy is now pushed to the PCM that is placed between a metal stripe and the external electrode in the vertical photodetector arrangement (Fig. 5 and 6). The metal stripe and the external electrode with the PCM inside form the MIM plasmonic waveguide that is characterized by high electric field enhancement in the gap [9, 23], *i.e.*, here in the PCM. As a result, it provides further temperature increases of the PCM that can be monitored by the electrical circuit.

In case of the lateral arrangement (Fig. 4), a heating of the PCM by the optical pulses cause the propagating mode is force to the metal stripe-PCM interface that is characterized by higher absorption in metal compared to the LR-DLSPP mode.

**Operation modes**

Under an efficient heating of the PCM by the light (optical pulses) the phase of PCM can be transferred from the amorphous to the crystalline. However, this approach will need very high amount of incident light on the PCM layer to make such a transition.

To enhance transfer process the electrical heating can be involved. Thus, under an applied voltage to the electrodes the heat will dissipated to the PCM causing the temperature increases of the PCM. When the voltage is close to the threshold voltage, the temperature in PCM is close to the crystallization temperature. As a result, the optical pulses can provide an additional heat source just to slightly farther increases the PCM temperature that can exceed the crystallization temperature. In consequence, the PCM undergoes the phase transition from an amorphous to crystalline state. The change of phase of PCM can be detected by the electrical circuit as the PCM transforms from the high resistance state to the low resistance state. Simultaneously, a transition of PCM can be detected through the optical measurements as an optical transmission of light changes from a high to a low transmission level. Inspired by Sarwat work [47], two operation modes can be considered.

Firstly, the proposed photodetector can be designed to operate in a count rate mode [47], in which a light coupled to the photodetector can be inferred from the rate at which the readout circuit resets the detector.

In such case, a bias voltage applied to the electrical circuit is used to initially heat up the PCM material, which is initially in an amorphous state and is a part of the LR-DLSPP-based photodetector, to the temperature that is close to the crystallization temperature $T_c$ of the PCM. In a temperature above $T_c$, the PCM undergoes a transition from an amorphous to a crystalline state. Thus, a relatively small amount of incident light will result in an increase in temperature beyond $T_c$, resulting in a phase transition of the PCM to a crystalline state. In consequence, a light coupled to the photodetector serves as an additional heat source just to provide a transition of the PCM from an amorphous to a crystalline state of PCM. Thus, under an applied voltage, at the initial stage, the photodetector operates under a high resistance state. When a light couple to the photodetector, the photodetector undergoes a transition to a crystalline state that is characterize by a low resistance state. It results in an increase in a current flow through PCM which can be detected by the readout circuit. After that, the readout circuit can reset the photodetector by applying the bias voltage to further increases the PCM temperature about its melting point $T_m$. In consequence, the PCM layer is re-amorphized and is ready for another pulses of light. The re-amorphization process is very fast and can be on the scale of nanoseconds.

The sensitivity of the photodetector can be adjusted by a bias voltage. For a high bias voltage, the temperature of the PCM is very close to the crystallization temperature, thus a low amount of light or low amount of optical pulses of light are needed to move PCM to the crystallization state.

Apart from the electrical readout circuit, a transition of PCM under a light coupled to the photodetector can be detected through the optical measurements. In the optical operation mode, an optical transmission of light under a transition of PCM from an amorphous to a crystalline state changes from a high transmission to a low transmission level.

As generally the re-amorphization time is faster than the crystallization time [20], the photodetector can operate in high bias voltage to initially heat the PCM to a temperature that is close to its melting temperature $T_m$. Thus, a photodetector can operate at the crystalline state under a high bias voltage. Then, when a light couple to the photodetector, the PCM is additionally heated by light and the temperature of PCM rises above $T_m$. In consequence, the PCM undergoes a fast transition from a crystalline state to an amorphous state. The change of state of PCM can be monitored by a readout circuit showing a change from a low resistance state to a high resistance state and/or by an optical transmission measurement. Depending on the material choice of the PCM and a design of the LR-DLSPP waveguide, the photodetector can operate in a low optical transmission level for a small amount of optical pulses, low level of laser flux, when a photodetector initially stays in a crystalline state of PCM.

In a second operation mode, the proposed photodetector can operate in a sub-threshold mode [17], in which a photodetector operates only in one state of the PCM, either an amorphous or a crystalline. In this operation mode, there is no transition of phase of PCM under a light coupled to the photodetector. Thus, a light coupled to a photodetector can be inferred either from a resistance of the phase change material (in the amorphous or crystalline state) measured by a readout circuit or from a transmission of light through a photodetector. In this operation mode, a reset of the photodetector can occur only when an unusually high rate of light is incident on the photodetector.

The proposed photodetector can employ both a vertical and lateral structure. In a vertical structure the PCM layer is sandwiched between a metal stripe and an external electrode while in a lateral structure the external electrodes are onto or into the PCM layer on both side of a ridge. Thus, a lateral structure is configured to pass current through the PCM layer laterally. The photodetector can comprise a further encapsulation layer (not shown) to protect the PCM from degradation.

The photodetectors, as shown in Fig. 4, 5, and 6, contain a readout circuit comprises a voltage source $V$, bias voltage $V_b$ and current photodetector $A$. A voltage source can be a programmable voltage source that is electrically connected to the metal stripe and the external electrode (Fig. 5 and 6) or to the first and second external electrode (Fig. 4). A bias voltage $V_b$ is applied to the PCM layer via the metal stripe and external electrode (vertical arrangement) (Fig. 5 and 6) or via the first and second external electrode (lateral arrangement) (Fig. 4). A current detector $A$ is implemented in a presented design to monitor the amount of current flowing through the PCM layer as a result of the applied bias voltage.

Under an absorption of light coupled to the photodetector by a phase change material, the PCM undergoes a transition from an amorphous to a crystalline phase. As the crystalline phase is characterize by a low resistance and the amorphous phase by a high resistance, the amount of light coupled to a photodetector can be monitored as a change in the current flowing through the phase change material under the bias voltage.

Compared to the free space photodetectors that require at least partially transparent electrodes to couple a light to the absorbing PCM later [17], the proposed waveguide-integrated photodetector provides much higher flexibility in choice of the materials for electrodes. Thus, the electrodes can be made from gold, silver, copper, aluminum, tantalum, graphene, transparent conductive oxides (TCOs), transition metal nitrides (TMNs), and *etc*. Furthermore, the metal stripe and the external electrode(s) can be from different materials.

The metal stripe and external electrode are configured to connect to a readout circuit in the vertical arrangement while a first external electrode and a second external electrode in the lateral arrangement.

**Activation function**

As the von Neumann architecture starts to fade with fast growing demands of the computing industry, the neuromorphic photonics became very promising field in the area of neuromorphic platforms with the purpose to mimic a behavior of human brain [14, 15, 16, 17, 20]. Furthermore, the neuromorphic photonics, or in the next stage the neuromorphic plasmonics, promise to overcome the electronic counterpart with lower power dissipation, higher footprint efficiency and faster throughput [14] (Fig. 7).

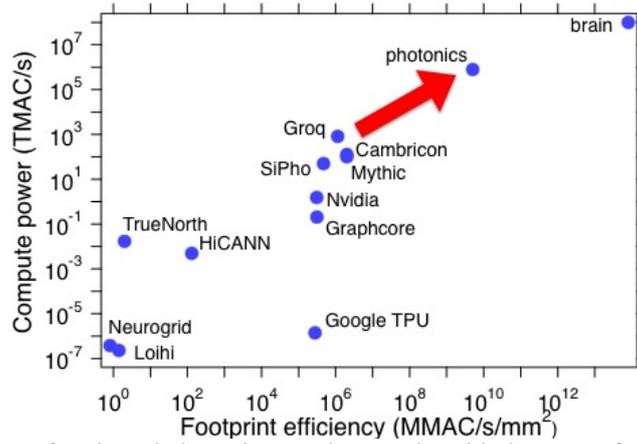

**Figure 7**. Comparison of projected photonic neural networks with the state-of-the-art electronic machines in terms of footprint efficiency and compute power.

Each artificial neuron in a network can be considered as two functional blocks containing a weighted addition unit and a nonlinear unit. The weighted addition, $\Sigma w_i x_i$, has multiple $x_i$ inputs that transmit the information to the neuron through the weights $w_i$, thus one output represents a linear combination of the inputs. It means that an optical carrier or a modulated signal $x_i$ needs to be attenuated of delayed by a certain amount dictated by a weight value $w_i$. The nonlinear unit applies an activation function, threshold, $f$ to the weighted sum to provide a nonlinear response, yielding the output of the neuron that is transmitted to many other neurons (Fig. 8a).

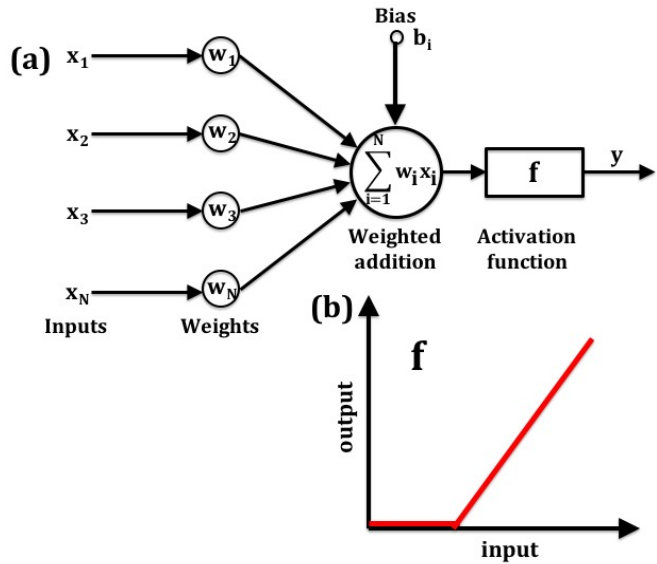

**Figure 8.** (**a**) Abstract representation of an artificial neuron with the weighted addition and the activation function. Fundamental operations in an artificial neuron: the weighted addition (or synaptic operation) and the activation (threshold) function. (**b**) Activation function required to achieve a nonvolatile phase transition.

The activation function mimics the firing feature of biological neurons. Thus, the nonlinear activation function can be used to set a threshold from which to define activated and deactivated behavior in artificial neurons [16, 46, 50] (Fig. 8b).
Incorporating optical nonlinearities into photonic circuits is one of the key requirements for deep photonic networks. One of the main challenges in implementing photonic nonlinear activation functions are due to the fact they need to operate at low optical signal intensities and repeat operations many times over [49, 50].
Nonlinear activation function elements can be classified in optoelectronic (OE) and all-optical (AO) ones [14, 16, 50]. The devices that are based on an optoelectronic (OE) nonlinear activation function unit need to first convert an optical signal into an electrical signal and afterwards convert it back into an optical signal. However, such mechanism highly impedes the speed and cascadability of the network [16]. Saturable absorbers represent all-optical (AO) solution where the amount of light absorbed decreases with increasing light intensity. They suffer however, from quick decay of the signal [15]. As showed in Ref. 15, if the transmissivity of the nonlinear activation function is $\sim\eta$ ($\leq 1$), then an optical signal with power $P_0$ propagating through an $L$-layer network with this activation function will decay exponentially to $\sim\eta^L P_0$.

Recently, it has been showed that electro-optic absorption modulators can be implemented for a nonlinear modulation of the optical signal [46]. The optical signal can be attenuated by the voltage that is able to vary the optical mode effective index and thus, a transmission of light.

The proposed arrangement can be designed to pass the optical pulses when the activation threshold is achieved and attenuate it when they are below the activation threshold. Thus, apart from a photodetection schema, the proposed device can be implemented as an activation function unit, meaning that a modulated signal can be attenuated or even delayed by a certain amount dictated by signals coming to the unit.

To realize such an activation function unit, the proposed arrangement can be designed to operate in a high-loss regime in the absence of the optical pulses that can be switched to a low-loss regime when the energy of the optical pulses coming to the unit exceeds the activation threshold. A low-loss regime corresponds the state in which the mode effective index below a metal stripe is equal or close to the mode effective index above a metal tripe. In such a situation, the absorption in metal stripe is minimalized and mode is characterized by a long propagation length.

In an opposite situation, *i.e.*, when the mode effective index below metal stripe is higher or lower than the mode effective index above a metal stripe, the absorption in metal stripe arise and propagation length drops down. The mode is pushed to one interface of metal stripe that is characterized by a higher mode effective index. As a distance between a metal stripe and an external electrode can be in a range of from tens to hundreds of nanometers, the mode propagating on one side of a metal stripe can couple to the external electrode, thus creating a gap surface plasmon polariton mode propagating in the metal-PCM-metal structure.

Compared to the LR-DLSPP mode that is characterized by a long propagation distance in the range of hundreds of micrometers, the MIM mode propagation distance usually no exceed few micrometers. For example, a propagation length of the Au-Si-Au structure with a Si gap width of 100 nm, was calculated at only 3 μm [48].

Thus, under a phase transition of the PCM, a huge difference in a propagation length of the plasmonic mode can be observed ranging from a few micrometers to up to hundreds of micrometers. In comparison, a previously reported MIM plasmonic waveguide with the GST placed in the gap showed a change in a light transmission lower than 4 % at telecom wavelength of 1550 nm under a transition of GST from an amorphous to a crystalline state [23]. In this case, under heating of the PCM a phase transition from an amorphous to a crystalline state occurs and the mode effective index of MIM increases. As a result, the mode is push more to the metals. Furthermore, the imaginary part of the GST highly increases in a crystalline phase what enhances the absorption losses. Thus, a farther reduction in a propagation length is observed under a transition of PCM, however this change is low.

The balance conditions in the mode effective indices can be monitored if an active material, such as for example PCM, is placed either in a buffer layer or in a ridge. Under, for example, a heating approach the refractive index of the PCM can be changed through a transition of the PCM between an amorphous and crystalline state.

Based on this approach, two different modes of operation of the proposed activation function unit can be distinguished.

Firstly, the activation function unit can be designed in a high-loss regime in an initial stage in the absence of the optical pulses, with the PCM layer in the crystalline state. To move the PCM to a crystalline state the electrical heating by the readout circuit is provided. The optical pulses coming to an activation function unit provide an additional heat source to the PCM layer. When the energy of the optical pulses exceeds an activation threshold, for which the temperature of the PCM exceeds the melting temperature of the PCM, the PCM undergoes a transition from a crystalline state to the amorphous state. As a mode effective index of a metal stripe and PCM in an amorphous state located either in a buffer layer (Fig. 4 and 5) or in a ridge (Fig. 6) is close to the mode effective index of the opposite side of the metal stripe, the balance in the mode effective indices is achieved that is characterized by low absorption losses in a metal stripe. In consequence, the propagation length reaches maximum and an activation function unit can operate in a low-loss regime.

The activation threshold of the activation function unit can be adjusted by a bias voltage from a readout circuit. For a high bias voltage, the temperature of the PCM can be very close to the melting temperature, thus a low amount of light or low amount of pulses of light can be needed to move PCM to the amorphous state.

In a second operation mode, the activation function unit can be designed in a high-loss regime in an initial stage in the absence of the optical pulses, however now, with the PCM layer in the amorphous state. In this scenario, the optical pulses can provide an efficient heat source to increases a temperature of the PCM above a crystallization temperature and transit a PCM to a crystalline state in which a balance conditions in the mode effective indices are established. As in a previous mode of operation, the activation threshold of the activation function unit can be adjusted by a bias voltage. An energy or the amount of optical pulses able to perform a transition from an amorphous to a crystalline state can be lowered through an initial heating of the PCM by the electrical bias voltage. To reset an activation function unit, a bias voltage for which the PCM transit from a crystalline state back to an amorphous state is required.

The proposed photodetector and activation function unit can be implemented into a recently proposed nonvolatile crossbar array as there are both on the same low-loss plasmonic waveguide platform and PCM [18]. Thus, a linear and nonlinear operations are performed via the same material class hence reducing fabrication complexity and system cost.

**Conclusion**

In summary, a low loss programmable and waveguide-integrated plasmonic photodetectors and activation function units that are based on PCMs were proposed and investigated. Different arrangements and heating configurations were considered to provide a most efficient mechanism for a detection and thresholding. It has been shown that the same device can operate either as a photodetector or an activation function unit depending on the operation conditions. Thus, it provides a flexibility in an on-chip signal processing. Furthermore, the proposed devices can provide a dual-mode operation in electrical and optical domains without a need of electro-optic conversion.

**Author information**
**Affiliations**
Independent Researcher, 90-132 Lodz, Poland
Jacek Gosciniak
**Contributions**
J.G. conceived the idea, performed all calculations and FEM and FDTD simulations and wrote the article.
**Corresponding author**
Correspondence to Jacek Gosciniak (jeckug10@yahoo.com.sg)